\documentclass[aip,apl,reprint]{revtex4-1}
\usepackage{graphicx}
\usepackage{dcolumn}
\usepackage{bm}

\usepackage[utf8]{inputenc}
\usepackage[T1]{fontenc}
\usepackage{mathptmx}
\usepackage{etoolbox}
\usepackage{siunitx}

\usepackage{hyperref}

\begin{document}

\preprint{AIP/123-QED}

\title{Phase-dependent parametric amplification of propagating spin waves in YIG nanostructures enabled by local inhomogeneities} 

\author{Akira Lentfert}
\email{lentfert@rptu.de}
\author{Ephraim Spindler}
\author{Björn Heinz}
\author{Mathias Weiler}
\author{Philipp Pirro}
 \homepage{https://pirro.physik.rptu.de/.}
\affiliation{ 
Department of Physics and State Research Center OPTIMAS, RPTU Kaiserslautern-Landau, 67663 Kaiserslautern, Germany
}%

\date{\today}
\begin{abstract}
As magnonics evolves towards non-conventional computing, the development of phase-conserving and phase-sensitive amplification mechanisms becomes increasingly important. A particularly promising approach is non-adiabatic parametric amplification. In this work, the influence of local inhomogeneities on the parallel parametric amplification of spin waves in nano-scale Yttrium Iron Garnet waveguides is investigated. Micromagnetic simulations reveal that in larger pump regions, where only adiabatic amplification is expected, scattering centers provide additional linear momentum that enables non-adiabatic amplification of propagating spin waves. Importantly, the coherence of the process remains unaffected by the scattering and the generation of co-propagating spin waves enhance the effective amplification. Our simulations are confirmed by micro-focused Brillouin light scattering spectroscopy experiments, reproducing both the phase-dependent behavior and the characteristic features of the time-resolved dynamics. These findings demonstrate the flexibility of the parametric amplification process and provide a key mechanism for the development of large-scale spin-wave computing circuits.
\end{abstract}

\maketitle 



In magnonics, the wave character of spin waves providing phase and amplitude as a degree of freedom has driven the development of logic elements\cite{Schneider2008, Chumak2014,Fischer2017, Wang2020a} forward. More recently, research has shifted towards non-conventional computing \cite{Flebus2024} schemes where the nonlinearity \cite{Merbouche2022, Wang2023} of spin waves is a crucial property, since concepts such as neuromorphic computing \cite{Papp2021} and reservoir computing \cite{Nakane2023,Watt2020} rely on power dependent spin-wave dynamics caused by nonlinear interactions.

In addition, for such large-scale magnonic networks, amplifiers are essential to re-amplify and normalize propagating spin waves between computing steps. Importantly, these amplifiers should maintain the phase coherency, have minimal noise figure, be unconditionally stable and emit spin waves only in the presence of an input to ensure the signal integrity.

Various amplification mechanisms, such as spin-transfer torque (STT) \cite{Merbouche2024} or stimulated amplification via rapid cooling \cite{Breitbach2023} have been explored. However, they are limited due to their interfacial nature (STT) or complex time dynamics (rapid cooling). Recently, an all-magnonic repeater \cite{Wang2024} based on bistable switching has been demonstrated that restores the spin-wave amplitude but simultaneously resets its phase, limiting its applicability for phase-based approaches.

A particularly promising candidate for phase-conserving amplification is parallel parametric pumping \cite{Bracher2014, Bracher2014a,Bracher2017a, Verba2018a, Nezu2024, Nikolaev2025}, as it solely requires a driving field oscillating at twice the spin-wave frequency. Here, parallel parametric generation refers to the amplification of thermal spin waves, whereas parallel parametric amplification describes the stimulated amplification of externally injected spin waves. The driving field is not restricted to a microwave field. Successful implementation using voltage-controlled magnetic anisotropy \cite{Verba2016} or surface acoustic waves in a magneto-elastic system \cite{Chowdhury2015,Geilen2025, Jander2025,Rivard2025a} has been demonstrated. Due to its flexibility, parallel parametric generation is widely employed for the mode-selective generation of spin waves in various structures, such as nano-waveguides \cite{Heinz2022, Hwang2021} and microdots \cite{Ulrichs2011, Hioki2021,Srivastava2023,Kim2024,Merbouche2024, Soares2026}.

The amplification process is classified as either adiabatic or non-adiabatic amplification and for the former case, phase-conservation and large amplification in nanostructures have already been realized\cite{Nikolaev2025}. Non-adiabatic amplification, by contrast, is more efficient and also phase-sensitive \cite{Melkov2001}, which is crucial for computing methods that encode the information in both phase and amplitude and rely on the coherency of the process. It has been successfully employed as a phase-to-intensity converter for the read-out of the phase information \cite{Bracher2016} but so far, is restricted to pumping fields localized to within half the spin-wave wavelength.

In this work, we show that the presence of local scattering centers in the form of material inhomogeneities allows for \textit{phase-dependent}, non-adiabatic parametric amplification of propagating spin waves using amplifier structures larger than half the spin-wave wavelength. We investigate the influence of inhomogeneities in a transversely magnetized, nanoscopic YIG waveguide of \SI{200}{\nano\metre} width by comparing micromagnetic simulations and micro-focused Brillouin light scattering (BLS) spectroscopy measurements. Our results contribute to the understanding of coherent amplification mechanisms in such nanoscale magnonic systems and increase the flexibility of the development of future programmable, phase-sensitive magnonic circuits.


Parallel parametric pumping describes the process where one microwave photon decays into two spin wave quanta (magnons) when the parametric threshold is overcome while conserving energy and momentum \cite{Lvov1994,Melkov2001}. The amplification is classified as either adiabatic or non-adiabatic amplification. In the adiabatic case, two spin waves with opposite momentum/wave vector are generated and the amplification is phase independent.
In case of non-adiabatic amplification, the linear momentum $k_\mathrm{p}=2\pi/L_\mathrm{p}$ provided by the localized pump antenna with a length $L_\mathrm{p}$ is larger than the sum of the signal and idler spin-wave wave vectors $k_\mathrm{s}$ and $k_\mathrm{i}$, allowing for the excitation of co-propagating spin waves. This minimizes the energy loss from back-propagating idler waves and prevents the interference with the incoming signal. The non-adiabaticity condition resulting from the momentum conservation is
\begin{equation}
    |k_\mathrm{s}+k_\mathrm{i}|<k_\textnormal{p} ,
    \label{eq:nonad_k}
\end{equation}
which is equivalent in case of $k_\mathrm{s} = k_\mathrm{i}$ to the condition
\begin{equation}
    \lambda_\textnormal{sw}>2L_\textnormal{p}.
    \label{eq:nonad_lambda}
\end{equation}

Here, we propose an alternative approach such that the non-adiabaticity condition is fulfilled without limiting the structure size of the pump antenna, allowing for a more flexible design of magnonic circuits. A local scattering center in form of a inhomogeneity of the width $w_\mathrm{inh}$ inside the pump area can provide an additional linear momentum $k_\mathrm{inh}=2\pi/w_\mathrm{inh}$, depending on its size. This adapts the wave vector condition Eq.~(\ref{eq:nonad_k}) to 
\begin{equation}
    |k_\mathrm{s}+k_\mathrm{i}|<k_\mathrm{p}+k_\mathrm{inh}, 
    \label{eq:nonad_defect}
\end{equation}
so that a smaller contribution from the pump antenna is sufficient for non-adiabatic amplification, maintaining the advantage of the phase-sensitivity.

In parallel parametric amplification, the sum of the phases of the generated, co-propagating spin-wave pair $\phi_\mathrm{s}$ and $\phi_\mathrm{i}$ fulfill the condition \cite{Lvov1994}
\begin{equation} \label{eq:pump_phase}
    \phi_\mathrm{s}+\phi_\mathrm{i}=\phi_\textnormal{p}^0+\pi/2.
\end{equation} 
for a pumping field with a fixed pump phase $\phi_\textnormal{p}^0$.
Since $\phi_\mathrm{s}$ is set by the phase of the incoming signal spin wave, the phase of the resulting idler wave $\phi_\textnormal{i}$ adjusts to fulfill Eq.~\ref{eq:pump_phase}. This results in a phase difference $\Delta\phi=|\phi_\mathrm{s}-\phi_\mathrm{i}|$ between the two co-propagating waves.
Due to the interference between them, this results in the amplification being $\propto\cos(\Delta\phi)$, with the largest amplification achieved for $\Delta\phi=0$.

For the investigation of the parallel parametric amplification process, in this work, a \SI{200}{\nano\metre} wide and \SI{112}{\nano\metre} thick YIG waveguide is structured from a full YIG film, grown on a \SI{500}{\micro\metre} thick gadolinium gallium garnet substrate by liquid phase epitaxy. 
\begin{figure}[h!]
    \centering
    \includegraphics[width=\linewidth]{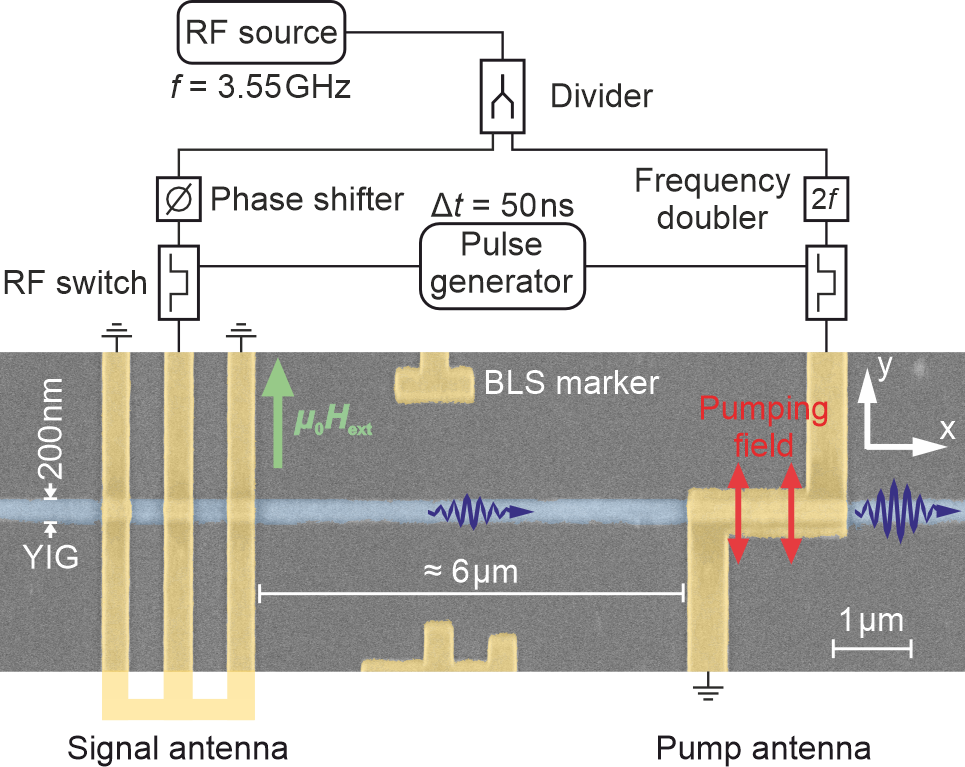}
    \caption{SEM micrograph image and schematic setup of the investigated structure (colored). All experiments were conducted for an external magnetic field of \SI{95}{\milli\tesla}, applied perpendicular to the YIG waveguide. Spin waves, excited by the signal antenna at x=0 propagate towards the pump antenna, where a delayed pump pulse allows for the parametric amplification of the signal spin waves.}
    \label{fig:structure}
\end{figure}
The magnetic parameters of the full film before structuring (${M_\textnormal{sat}= \SI{155}{\kilo\ampere\per\metre}}$ and  Gilbert damping parameter of ${\alpha=3.5\times10^{-4}}$) are extracted using ferromagnetic resonance (FMR) spectroscopy \cite{Harder2011}. 

Two types of titanium/gold antenna structures are patterned on top (see  Fig.~\ref{fig:structure}) of the waveguide, with a thickness of \SI{10}{\nano\metre}/\SI{140}{\nano\metre}. The first is a coplanar waveguide (CPW) which is referred to as "signal antenna", for which the first maximum of the excitation efficiency lies at a wave vector of \SI{3.11}{\radian\per\micro\metre}. The second antenna, referred to as "pump antenna", is placed about \SI{6}{\micro\metre} away and generates a pumping field with an effective width of \SI{1.41}{\micro\metre}. The central part of the pump antenna covers the waveguide completely, allowing for the generation of an oscillating magnetic field perpendicular to the waveguide. The external magnetic field is also applied perpendicular to the waveguide (Damon-Eshbach (DE) geometry). This geometry is advantageous for extended magnonic circuits, as the non-reciprocal excitation efficiency suppresses the back-propagating signals \cite{Jamali2013,Heinz2021} and the chiral nature of the spin waves reduces back-scattering from defects \cite{Mohseni2019}, both maintaining the signal integrity.

To obtain a spatially resolved spectrum of the excited spin waves, micro-focused BLS spectroscopy is employed \cite{Sebastian2015}. A laser with a wavelength of $\lambda=\SI{457}{\nano\metre}$ and a laser power of \SI{3}{\milli\watt} is focused on the structure using a microscope objective. This setup enables the detection of spin waves with a spatial resolution of approximately \SI{300}{\nano\metre}.

We perform micromagnetic simulations using the simulation package Mumax3 \cite{Vansteenkiste2014} and the software platform Aithericon \cite{Aithericon} to estimate the influence of a local inhomogeneity inside the pump area on the amplification process.
For an external magnetic field of \SI{95}{\milli\tesla}, a signal spin-wave pulse with a wave vector of \SI{3.05}{\radian\per\micro\metre} is launched and propagates into the area of the pumping field. The field distribution of the pump antenna was derived using a full-wave planar electromagnetic field solver (Sonnet Suites \cite{Sonnet}). The pumping field is switched on with a delay of \SI{50}{\nano\second}, such that the signal spin waves already arrived in the pump area and the resulting time-dependent out-of-plane magnetization along the center of the conduit is extracted. To avoid the additional generation of spin-waves unrelated to the incoming signal, the strength of the pumping field is limited to values below the generation threshold field \cite{Bracher2014}. For our chosen pump antenna, only adiabatic amplification is expected for the ideal case. More details on the simulation parameters can be found in the supplementary material.

In Fig.~\ref{fig:Adiabatic_anisotropy_phasecolorplot}~a), the normalized amplitude of the spin-wave pulse as a function of the phase difference $\Delta\phi$ in the inhomogeneity-free waveguide is shown. $\Delta\phi$ is well defined according to Eq.~\ref{eq:pump_phase} by the phase of the signal spin waves $\phi_\mathrm{s}$, which is varied in the simulations, and the fixed pump phase.
\begin{figure}[h!]
    \centering
    \includegraphics[width=\linewidth]{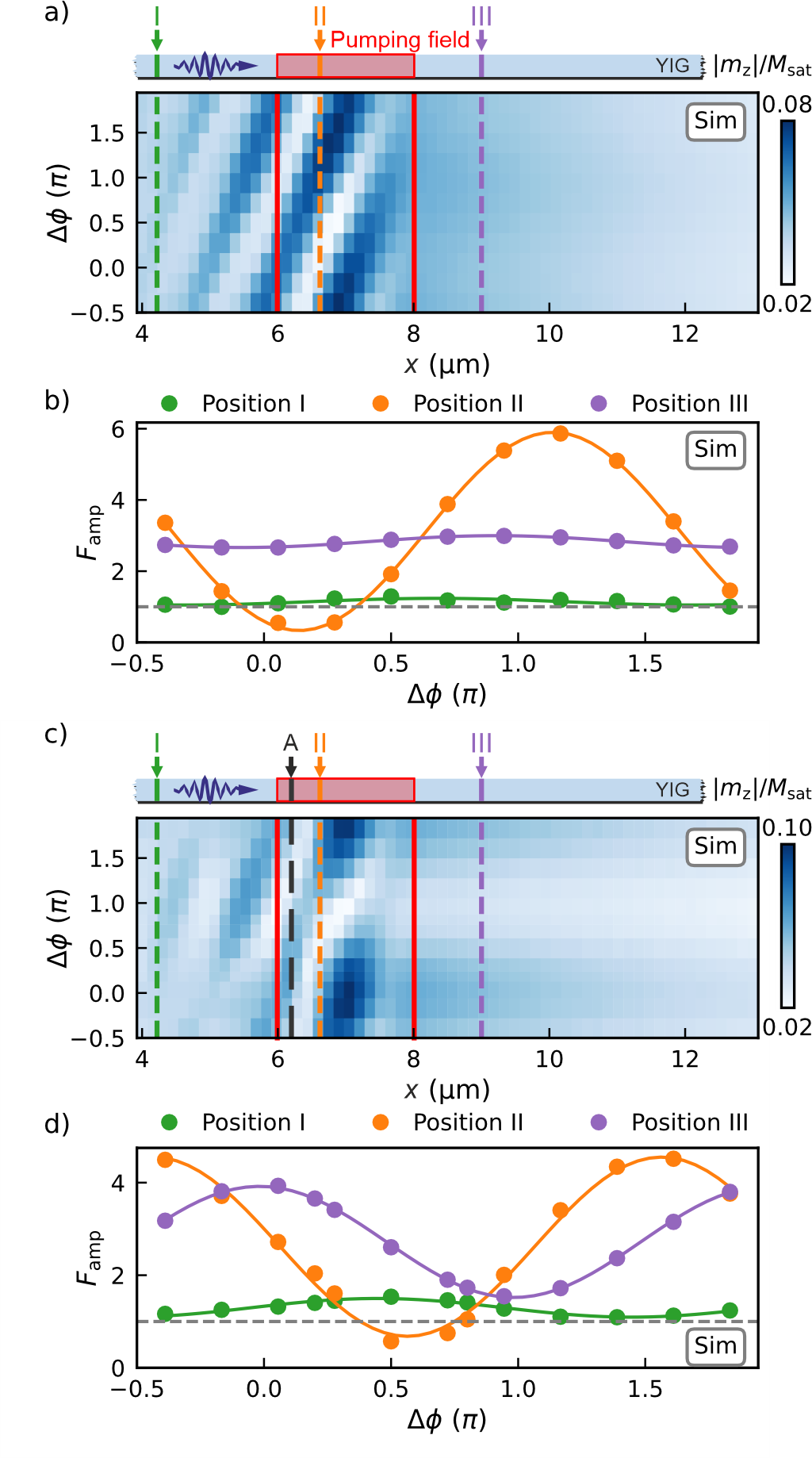}
    \caption{Micromagnetic simulations of the phase dependence in a), b) an ideal waveguide and c), d) a waveguide with a inhomogeneity at position A at $x=\SI{6.2}{\micro\metre}$. In a), c) the normalized amplitude of the spin-wave pulse extracted from micromagnetic simulations is shown as a function of the phase difference $\Delta\phi$ and the position along the waveguide. The pumping field is applied in the region between the red vertical lines. b), d) The amplification factor, extracted for the highlighted distinct positions are shown.}
    \label{fig:Adiabatic_anisotropy_phasecolorplot}
\end{figure}
The interference of forward-propagating signal spin waves and backward-propagating idler waves generated during the amplification strongly depends on the phase difference $\Delta\phi$ between them. In Fig.~\ref{fig:Adiabatic_anisotropy_phasecolorplot}~b), the extracted amplification factor $F_\mathrm{amp}$ is shown for three positions along the waveguide. $F_\mathrm{amp}$ is defined as the ratio between the maximum intensity of the amplified spin-wave pulse and the maximum intensity of the sum of the two individual pulses at the same position: 
\begin{equation} \label{eq:famp_definition}
    F_\mathrm{amp}(\Delta\phi)=\mathrm{max}\{I(t,\Delta\phi)\}/\mathrm{max}\{I_\mathrm{s}(t)+I_\mathrm{p}(t)\}
\end{equation}
Behind the pump antenna, no phase dependence appears, showing that the ideal system indeed allows for adiabatic amplification.

To estimate the influence of a local inhomogeneity, perpendicular magnetic anisotropy is added to a $w_\mathrm{inh}=\SI{200}{\nano\metre}$ wide area at position \SI{6.2}{\micro\metre}, homogeneous over the waveguide width. The uniaxial anisotropy constant is set to ${K_\textnormal{u1}=\SI{200}{\joule\per\cubic\metre}}$, corresponding to an additional anisotropy field of \SI{3.7}{\milli\tesla}, well below three percent of the saturation magnetization contribution.

Figure~\ref{fig:Adiabatic_anisotropy_phasecolorplot}~c) shows the resulting phase dependence including the local inhomogeneity. Now, a clear change is visible behind the pump antenna, indicating a transition to a non-adiabatic amplification process. It is important to mention that the change of the effective dispersion relation inside the local inhomogeneity is not critical to obtain this result. Instead, one can interpret the local inhomogeneity as a scattering center, which provides an additional linear momentum $k_\mathrm{inh}$, depending on its size, to the amplification. This adapts the wave vector condition to Eq.~\ref{eq:nonad_defect}, allowing for a phase dependent amplification in larger pump structures. Similar results can be obtained when locally modifying other material parameters such as the saturation magnetization. This is shown in the supplementary materials.

The resulting amplification factor $F_\mathrm{amp}$ is shown as a function of the phase difference for the same three positions in Fig.~\ref{fig:Adiabatic_anisotropy_phasecolorplot}~d). For all cases, $F_\mathrm{amp}$ exhibits a $2\pi$ periodicity in $\Delta\phi$, as derived from the interference condition. The dependencies between position 2 and 3 are shifted due to the complex interference of forward and backward propagating waves inside the pump area. Compared to the adiabatic case (Fig.~\ref{fig:Adiabatic_anisotropy_phasecolorplot}~b)), larger amplifications are achieved behind the antenna, since both parametrically generated spin waves contribute to the signal. As a reference case, a defect free structure with a \SI{300}{\nm} wide pumping field is simulated, which naturally meets the condition for non-adiabatic amplification. Here, the same phase-dependence is observed, as shown in the supplementary material.

For further optimization of the amplification process, $F_\mathrm{amp}$ is determined for varying defect widths $w_\mathrm{inh}$ up to \SI{8}{\micro\metre}.
\begin{figure}
    \centering
    \includegraphics[width=\linewidth]{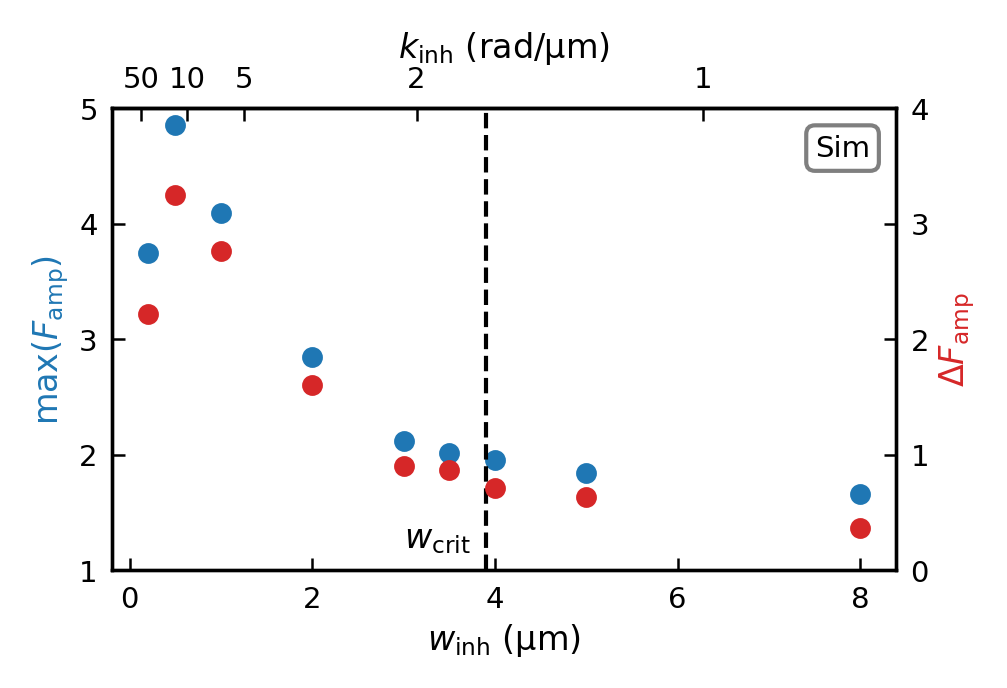}
    \caption{Simulated dependence of $\max(F_\mathrm{amp})$ and the amplitude of the modulation by the phase $\Delta F_\mathrm{amp}$ is shown as a function of the width $w_\mathrm{inh}$ and the maximum linear momentum $k_\mathrm{inh}$. In black, the transition between adiabatic and non-adiabatic amplification at $w_\mathrm{crit}$ is highlighted.}
    \label{fig:width_dep}
\end{figure}
For a square shaped inhomogeneity, as studied above, the provided linear momentum has a sinc-shaped distribution up to $k_\mathrm{inh}$. This means that in our system, inhomogeneities smaller than $w_\mathrm{crit}=\SI{3.9}{\micro\metre}$ should provide enough additional linear momentum to achieve non-adiabatic amplification. Figure~\ref{fig:width_dep} shows the maximum amplification factor $\max(F_\mathrm{amp})$ as a function of $w_\mathrm{inh}$. The positive impact of smaller inhomogeneity sizes on the amplification is clearly visible. Larger inhomogeneities lead to increased damping due to the changes of the dispersion inside, while when downsized too much, the interaction length becomes smaller. This conclusion can be drawn from the extracted modulation of the amplification $\Delta F_\mathrm{amp}=\max(F_\mathrm{amp})-\min(F_\mathrm{amp})$ too, which demonstrates how strong the dependence of the amplification on the phase difference $\Delta\phi$ is.

In the following, these micromagnetic results are experimentally validated by investigating the \SI{200}{\nano\metre} wide YIG waveguide shown in Fig.~\ref{fig:structure} using BLS. 
During the characterization of the nanostructure, a local inhomogeneity close to the boundary between the covered and uncovered YIG towards the signal antenna is identified. When investigating the parallel parametric generation of the pump antenna, spin waves propagating towards the $-x$ direction are strongly damped independent of the external field direction aligning in + or $-y$ direction, while propagatiopm along $+x$ remains unaffected (see supplemental material). Since adding metal structures on top can induce chemical modifications and elastic stress\cite{Wang2014}, which alter material parameters and for example also result in the pinning of domain walls \cite{Fan2023}, we attribute this finding to such a local defect in the structure.

To study the amplification mechanism, \SI{50}{\ns} long Rf pulses are applied to the signal and pump antennas, which are synchronized and phase locked. The driving frequency is \SI{3.55}{\giga\hertz} and the external magnetic field is fixed at \SI{95}{\milli\tesla}. Under these conditions, we excite spin waves at \SI{3.55}{\giga\hertz}. The signal spin-wave pulse first arrives inside the pumping area, followed by the pump pulse applied with a \SI{20}{\nano\second} delay. For the amplification experiments, a pump power slightly above the generation threshold was applied. More information concerning the influence of the delay on the amplification is presented in the supplementary material.

\begin{figure}[h!]
    \centering
    \includegraphics[width=\linewidth]{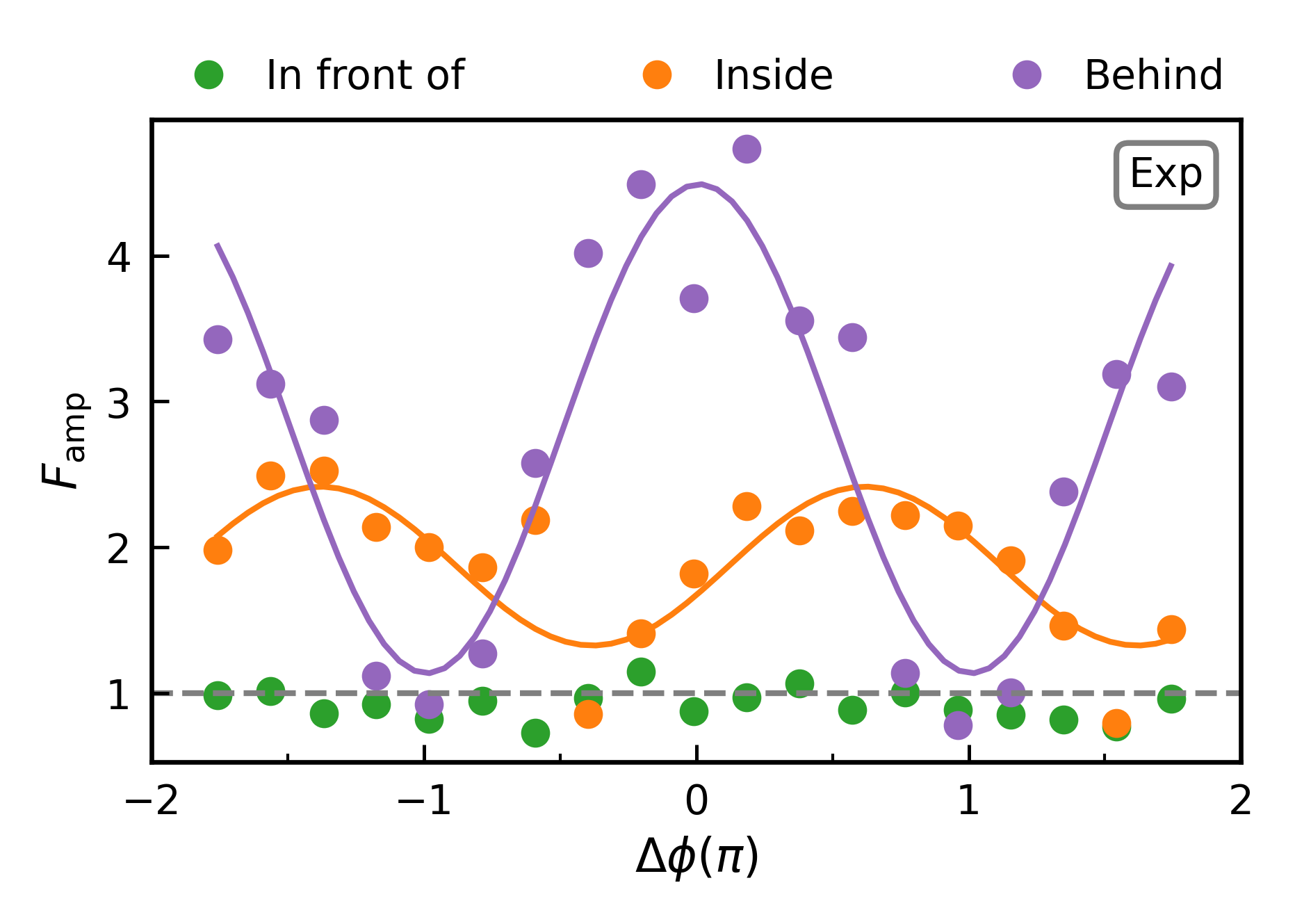}
    \caption{Measured phase dependence of the amplification at different positions. The amplification factor $F_\mathrm{amp}$, obtained by evaluating the pulse amplitudes according to \ref{eq:famp_definition}, for varying phase differences $\Delta\phi$ at the center of the pump antenna and \SI{1.5}{\micro\metre} behind and in front of it are shown.}
    \label{fig:Phase_dependent_amp_M139}
\end{figure}

In Fig.~\ref{fig:Phase_dependent_amp_M139}~a), $F_\mathrm{amp}$ extracted from the time-resolved data for all measured phase differences is shown for positions inside and \SI{1.5}{\micro\metre} in front of and behind the pumping area. 
In front of the pump antenna, no amplification is observed since back-propagating spin waves are strongly damped by the local inhomogeneity. 
In contrast, a clear phase dependence is observed inside and behind the pumping area, even surpassing $F_\mathrm{amp}=4$ and being in good agreement with the simulated results shown in Fig.~\ref{fig:Adiabatic_anisotropy_phasecolorplot}~d). The fit for the data at the position inside and behind shows the same, shifted $2\pi$ periodicity on $\Delta\phi$.

Based on the realization of non-adiabatic amplification described above, we now focus on the resulting spin wave pulse shape. A comparison with the simulations is performed by examinating the time-resolved data and furthermore, the compatibility towards electrical detection schemes is evaluated.

\begin{figure}[h!]
    \centering
    \includegraphics[width=\linewidth]{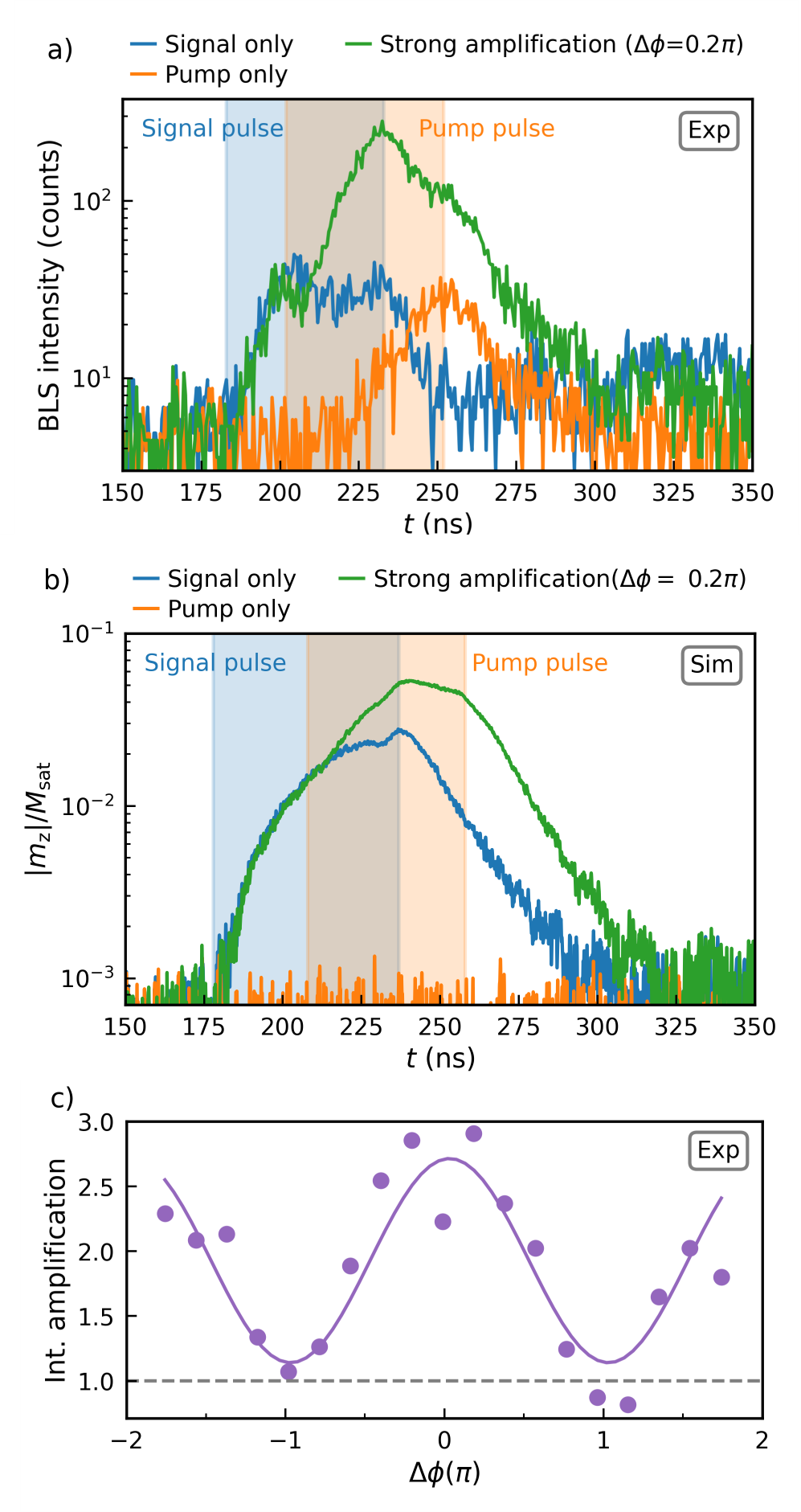}
    \caption{a) Time-resolved BLS measurements for strong amplification at $\Delta\phi=0.2\pi$ are shown. The shaded areas highlight the time intervals of the applied RF pulses. b) Simulated spin-wave pulses in case of the inhomogeneity enabled amplification process at position 3 (see Fig.~\ref{fig:Adiabatic_anisotropy_phasecolorplot}~c)) for the same phase difference $\Delta\phi$. c) The amplification ratio of the time-integrated signals from a) is plotted as a function of the phase difference $\Delta\phi$.}
    \label{fig:timetrace_nonad_M139}
\end{figure}

In Fig.~\ref{fig:timetrace_nonad_M139}~a), the amplified spin-wave pulses measured behind the pump antenna in case of the strong amplification at the phase difference $\Delta\phi=0.2\pi$ is shown. The time-dependent spin-wave intensity, when only the pumping source is switched on, is shown in orange. Since a pump power slightly above the generation threshold is applied, the intensity increases slightly towards the end of the pump pulse. 

For the present case, the spin wave pulse is amplified as long as both pulses overlap in time and the largest amplitude is achieved at the end of the signal pulse (green). The pump pulse alone cannot maintain the large spin-wave amplitude, resulting in the drop of the intensity as soon as the signal pulse leaves the pump area.

In the micromagnetic simulations in Fig.~\ref{fig:timetrace_nonad_M139}~b), the spin-wave intensity increases as soon as the total energy flux into the mode compensates the dissipation during the overlap interval. Similar to the experiments, largest amplitude is achieved at the end of the signal pulse, after which it begins to drop during the remaining time interval of the pump pulse. It should be noted that minor differences of the spin wave pulse shape can be attributed to the simplified modelling of the thermal spin wave distribution and the scattering center. While the chosen inhomogeneity allows for the accurate reproduction of the overall phase dependence, the influence on for example the damping and time evolution could deviate from the investigated structure.
A comparison of these simulations with the case of strictly non-adiabatic amplification, presented in the supplementary materials, demonstrates that the temporal behavior is comparable. This underlines the successful transition from adiabatic to non-adiabatic amplification enabled by local inhomogeneities.

When working with electrical detection schemes, the pulse energy and therefore the shape of the pulses becomes important in addition to their absolute amplitudes. For future applications, the electrical read-out of spin waves is of great interest for establishing a link to the fields of spintronic and electronics. This can be achieved, for example, by means of the inverse spin-Hall effect\cite{Kajiwara2010}. To determine the spin-wave pulse energy after amplification, the time-resolved data shown in Fig.~\ref{fig:timetrace_nonad_M139}~a) is integrated over time for each $\Delta\phi$-value. The results are normalized by taking the ratio to the sum of the signal and pump pulses. The resulting integrated amplification factor is shown in Fig.~\ref{fig:timetrace_nonad_M139}~c). The elongation of the spin-wave pulse by the amplification process leads to an amplification factor of up to three for the integrated signal, making it a promising candidate as a phase-dependent signal amplifier for electrical read-out.


In summary, we demonstrated that local inhomogeneities in nanostructured magnonic waveguides enable non-adiabatic parametric amplification for extended pump areas, achieving larger efficiencies than adiabatic pumping as both parametrically generated spin waves contribute to the signal. Complementary experiments confirmed that the presence of an inhomogeneity leads to the emergence of phase-dependent amplification with comparable performance. Our findings leave room for further optimizations through inhomogeneity patterning and larger pump antennas, advancing the development of large-scale, phase-sensitive magnonic networks for advanced information processing.
\begin{acknowledgments}
The authors acknowledge support by the European Research Council within the Starting Grant No. 101042439 ”CoSpiN”; the Deutsche Forschungsgemeinschaft (DFG, German Research Foundation) through the Transregional Collaborative Research Center-TRR 173-268565370 ‘Spin+X’ (Project B11), and by the Federal Ministry of Research, Technology and Space (BMFTR) under the reference number 13N17110. The authors acknowledge support by the Nano Structuring Center of RPTU Kaiserslautern-Landau.
\end{acknowledgments}


%
%

%


\bibliography{references}
\end{document}